\documentclass{aa}
\usepackage{graphicx}
\usepackage{txfonts}
\usepackage{hyperref}
\usepackage{natbib,url,twoopt}
\usepackage{amssymb,bm}
\usepackage{multirow}                           
\hypersetup{
  colorlinks=true,   
  urlcolor=blue,     
  linkcolor=red,     
}
\newcommand{\mas}{$\,\mathrm{mas}$}   
\newcommand{\masyr}{$\,\mathrm{mas\,yr^{-1}}$}   
\newcommand{\uas}{$\,\mathrm{\mu as}$} 
\newcommand{\uasyr}{$\,\mathrm{\mu as\,yr^{-1}}$} 
\newcommand{\kms}{$\,\mathrm{km\,s^{-1}}$} 
\newcommand{\kmskpc}{$\,\mathrm{km\,s^{-1}\,kpc^{-1}}$} 
\begin{document}

   \title{Overall properties of the {\it Gaia} DR1 reference frame}

   \author{N. Liu
          \and
          Z. Zhu
          \and
          J.-C. Liu
          \and
          C.-Y. Ding
          }

   \institute{School of Astronomy and Space Science, Key Laboratory of Modern Astronomy and
Astrophysics (Ministry of Education), Nanjing University, 163 Xianlin Avenue, 210023 Nanjing, P.~R.~China\\
              \email{zhuzi@nju.edu.cn}
             }

   \date{Received ; accepted }


  \abstract
   {}
   {The first {\it Gaia} data release ({\it Gaia} DR1) provides 2\,191 ICRF2 sources with their positions in the auxiliary quasar solution and
    five astrometric  parameters -- positions, parallaxes, and proper motions -- for stars in common between the {\it Tycho}-2 catalogue and {\it Gaia} in the joint {\it Tycho-Gaia} astrometric solution (TGAS). We aim to analyze the overall properties of {\it Gaia} DR1 reference frame.}
   {We compare quasar positions of the auxiliary quasar solution with ICRF2 sources using different samples and evaluate the influence on  the {\it Gaia} DR1 reference frame owing to the Galactic aberration effect over the J2000.0-J20015.0 period.
   Then we estimate the global rotation between TGAS with {\it Tycho}-2 proper motion systems to investigate the property of the {\it Gaia} DR1 reference frame.
   Finally, the Galactic kinematics analysis using the K-M giant proper motions is performed to understand the property of {\it Gaia} DR1 reference frame.}
   {The positional comparison between the auxiliary quasar solution and ICRF2 shows negligible orientation and validates the declination bias of $\sim$$-0.1$\mas~in {\it Gaia} quasar positions with respect to ICRF2.
Galactic aberration effect is thought to cause an offset $\sim$$0.01$\mas~of the $Z$ axis direction of {\it Gaia} DR1 reference frame.
The global rotation between TGAS and {\it Tycho}-2 proper motion systems, obtained by different samples, shows a much smaller value than the claimed value $0.24$\masyr.
For the Galactic kinematics analysis of the TGAS K-M giants, we find possible non-zero Galactic rotation components beyond the classical Oort constants: the rigid part $\omega_{Y_G} = -0.38 \pm 0.15$\masyr~and the differential part $\omega^\prime_{Y_G} = -0.29 \pm 0.19$\masyr~around the $Y_G$ axis of Galactic coordinates, which indicates possible residual rotation in {\it Gaia} DR1 reference frame or problems in the current Galactic kinematical model.}
   {The {\it Gaia} DR1 reference frame is well aligned to ICRF2, and the possible influence of the Galactic aberration effect should be taken into consideration for the future {\it Gaia}-ICRF link.
   The cause of the rather small global rotation between TGAS and {\it Tycho}-2 proper motion systems is unclear and needs further investigation.
   The possible residual rotation in {\it Gaia} DR1 reference frame inferred from the Galactic kinematic analysis should be noted to and examined in future data release.
   }

   \keywords{
                astrometry --
                        reference systems --
                        proper motions
               }

   \maketitle
%

\section{Introduction}     \label{sec:introduction}

The first {\it Gaia} data release \citep[{\it Gaia} DR1;][]{2016A&A...595A...2G} on 14 September 2016 provides astrometric parameters for nearly 2 million stars and positions for over 1 billion sources with improved accuracy, and leads to several improvements in fundamental astrometry, especially in the practical realization of the optical reference frame.

The definition and realization of the reference frame has been one of the main concerns for astrometry.
The {\it Gaia} reference frame links directly to the extragalactic objects through the quasar observations, and surpasses the former {\it Hipparcos} reference frame in accuracy and inertia.
The final {\it Gaia} celestial reference frame is supposed to reach a level of a few tens of micro-arcseconds ($\mathrm{\mu as}$), similar to the International Celestial Reference Frame (ICRF).
The {\it Gaia}-ICRF link will be important and has been used in the construction of the third generation of ICRF \citep[ICRF3; for example, see][]{2014adla.confE...1J}.
A detailed analysis of the {\it Gaia} reference frame and ICRF should be performed to prepare for the future optical-radio link.

As a preliminary result of the {\it Gaia} reference frame, the {\it Gaia} DR1 reference frame has been aligned to ICRF better than 0.1\mas~at epoch J2015.0 \citep{2016A&A...595A...4L}.
Analysis of the {\it Gaia} DR1 reference frame can help to illustrate potential properties of the final {\it Gaia} reference frame.
Detailed comparisons and analysis between {\it Gaia} DR1 with other catalogues have been performed in \citet{2016A&A...595A...4L} and \citet{2016A&A...595A...5M}, including the optical property of ICRF sources.

This work aims to provide an extra overall property analysis of the {\it Gaia} DR1 reference frame.
We begin with some tests of the auxiliary quasar solution in Sect.~\ref{sec:QSOcomparison}. Then the analysis of the the joint {\it Tycho-Gaia} astrometric solution (TGAS) proper motion system have been performed in light of global rotation (Sect.~\ref{sec:GlobalRotation}), and Galactic kinematics (Sect.~\ref{sec:KinematicsAnalysis}).
\section{Overall analysis of the radio reference frame}    \label{sec:QSOcomparison}
\subsection{Overall property of the auxiliary quasar solution}
The auxiliary quasar solution in {\it Gaia} DR1 was used to align the {\it Gaia} DR1 reference frame to the second generation of ICRF (ICRF2) by a solid rotation \citep[][Eq.~5]{2016A&A...595A...4L}
The available version contains 2\,191 ICRF2 sources with their positions in the {\it Gaia} DR1 reference frame.
\citet{2016A&A...595A...5M} provide a detailed comparison between the auxiliary quasar solution and the ICRF2 catalogue and find a systematical declination bias at the order of $-$0.1\,mas.

The declination bias can be confirmed and then removed by adding a parameter $\mathrm{d}z$ to the declination component of the Eq.\,(5) in \citet{2016A&A...595A...4L},  as performed in the alignment between ICRF1 and ICRF2 \citep[][]{2015AJ....150...58F}.
As such, the equation for alignment can be written as \citep[e.g.][]{2006A&A...452.1107F} :
\begin{equation}\label{eq:aligment}
   \begin{array}{lll}
\Delta\alpha ^* & = & -\epsilon_X\cos\alpha\sin\delta - \epsilon_Y\sin\alpha\sin\delta + \epsilon_Z\cos\delta \\
\Delta\delta    & = &+\epsilon_X\sin\alpha                      - \epsilon_Y\cos\alpha + \mathrm{d}z
   \end{array}
,\end{equation}
where $\Delta\alpha ^*=\Delta\alpha\cos\delta$.
Parameters $\epsilon_X$, $\epsilon_Y$, and $\epsilon_Z$ are three relative rotation angles between two celestial reference frames around the $X$, $Y,$ and $Z$ axes and $\mathrm{d}z$ accounts for the systematical declination differences, which is caused by the possible inaccuracy of the tropospherical model, in the case of alignment between ICRF1 and ICRF2.
The positional differences ($\Delta\alpha ^*$, $\Delta\delta$) between {\it Gaia} and ICRF2 is calculated  based on {\it Gaia}$-$ICRF2.
To verify the declination bias, we used four subsets: all 2\,191 sources, all 262 defining sources, 1\,929 non-defining sources, and 260 defining sources used for fixing the {\it Gaia} DR1 frame (two sources 0119+115 and 1823+568 are only used for the right ascension component).
For comparison, we also estimated just the solid rotation between the auxiliary quasar solution and the ICRF2.

\begin{table*}
\caption{\label{tab:orientation}
Orientations of the auxiliary quasar solution relative to ICRF2.}
\centering
\begin{tabular}{c c c c c c }
\hline \hline
Subset          &$N$            &  $\epsilon_X$         & $\epsilon_Y$           & $\epsilon_Z$ &$\mathrm{d}z$   
         \\ [0.5ex]
\hline
\multirow{2}{*}{All}                    &\multirow{2}{*}{2\,191}        
                                        &$+62 \pm 16$           &$+35 \pm 15$    &$+21 \pm 16$    &       \\
                        &               &$+75 \pm 16$           &$+49 \pm 15$    &$+18 \pm 16$    &$-123 \pm 13$ \\
\multirow{2}{*}{Defining}               &\multirow{2}{*}{262}                   
                                        &$+31 \pm 29$           &$+41 \pm 25$    &$ -36 \pm 32$   &       \\
                        &               &$+42 \pm 29$           &$+54 \pm 26$    &$ -39 \pm 32$   &$-78 \pm 23$ \\
\multirow{2}{*}{Non-defining}   &\multirow{2}{*}{1\,929}                                
                                        &$+79 \pm 20$           &$+31 \pm 18$    &$+39 \pm 18$    &       \\
                        &               &$+91 \pm 20$           &$+44 \pm 18$    &$+36 \pm 18$    &$-143 \pm 16$ \\
\multirow{2}{*}{Frame-fixed}    &\multirow{2}{*}{260}                   
                                        &$ - 0 \pm 29$          &$ -  0 \pm 25$    &$ - 0 \pm 32$   &       \\
                        &               &$+17 \pm 30$           &$+20 \pm 26$    &$ - 5 \pm 32$   &$-121 \pm 23$   \\             
\hline
\multirow{2}{*}{Northern}       &\multirow{2}{*}{1\,298}                                
                                        &$+131 \pm  20$     &$+ 72 \pm  19$    &$+102 \pm  21$      &       \\
                        &               &$+138 \pm  20$     &$+111 \pm  19$    &$+99 \pm  21$       &$-140 \pm  17$ \\
\multirow{2}{*}{Southern}       &\multirow{2}{*}{893}                   
                                        &$-84 \pm  28$          &$-26 \pm  23$    &$-85 \pm  24$  &       \\
                        &               &$-64 \pm  28$          &$-41 \pm  24$    &$-85 \pm  24$  &$-113 \pm  21$  \\[0.5ex]
\hline
\end{tabular}
\tablefoot{The unit is\uas.}
\end{table*}
From the results of the least squares using Eq.~(\ref{eq:aligment}) (Table~\ref{tab:orientation}), we can clearly see a declination bias ${\rm d}z \sim$$-0.1$\mas,~as reported in \citet{2016A&A...595A...5M}.
We note that the {\it Gaia} DR1 reference frame has been aligned to the ICRF2, and we should not expect any rotation components.
The results for mutual orientations from the full samples are insignificant and small compared to the claimed value 0.1~\mas~in \citet{2016A&A...595A...4L}, which meets expectations.

To obtain more insight of the systematic declination bias, we further only considered  the sources in the northern and southern hemispheres, respectively. 
The result is presented in the last two rows in Table~\ref{tab:orientation}.
On the one hand, both hemispheres yields a similar declination bias, which indicates that the systematic declination bias less possibly arises from the southern or northern declination offsets in the VLBI or {\it Gaia} data.
On the other hand, although the obtained result that no orientation component is larger than 0.15 mas agrees with that in \citet{2016A&A...595A...4L}, the signs of the orientation components from the northern and southern hemispheres are totally opposite, indicating possible regional deformations in the ICRF2 or {\it Gaia} DR1 reference frame.
Limited by the small sample available, it is hard to obtain a clear explanation so far.

\subsection{Galactic aberration effect on {\it Gaia} DR1 reference frame}
\citet{2016A&A...595A...5M} clearly claim  that no models of the Galactic acceleration have been introduced in auxiliary quasar solution.
The Galactic acceleration effect will be certainly removed in the final Gaia reference frame, but is not considered yet in present very long baseline interferometry (VLBI) data reductions.
A possible effect on the reference frame owing to the Galactic acceleration should be estimated.
In general, the Galactic acceleration of the solar system barycenter produces apparent proper motions for extragalactic sources, which results in a dipole pattern of apparent proper motions.
This is called the Galactic aberration effect \citep{2011ARep...55..810M}.
In addition, the celestial reference frame, based on the subset of extragalactic sources, has some systematic effects, such as global rotation and deformation, which results from the apparent proper motions.
Recently \citet{2016FrASS...3...28M} took the Galactic aberration effect into consideration and provided possible methods to create a future link of the {\it Gaia} reference frame with the next generation of ICRF.

We estimated the accumulated orientation difference from the global rotation due to the Galactic aberration effect over the J2000.0-J20015.0 period.
We note that the small orientation offset depends on the distribution of extragalactic sources \citep{2012A&A...548A..50L}.
Four subsets in Table~\ref{tab:orientation} are tested to evaluate the possible orientation offsets of the celestial reference frame owing to the Galactic aberration effect.
The orientation offset was denoted as three rotation angles $\Delta\epsilon_X$, $\Delta\epsilon_Y$, and $\Delta\epsilon_Z$ around the $X$, $Y,$ and $Z$ axes of the celestial reference frame.
We adopted the apex of the dipole pattern as the Galactic center and the Galactic aberration constant $A\,=\,5$\uasyr to obtain the apparent proper motions of extragalactic sources. Based on this, we then estimated the global rotation of the celestial reference frame.
Afterwards, the orientation offset was obtained by multiplying the global rotation by the 15-year epoch span (J\,2000.0--J\,2015.0).

\begin{table*}
\caption{\label{tab:GAEffect}
Orientation offset due to the Galactic aberration effect over the J2000.0-J20015.0 period.
}
\centering
\begin{tabular}{c c c c c }
\hline \hline
Subset          &$N$            &$\Delta\epsilon_X$             &$\Delta\epsilon_Y$                  &$\Delta\epsilon_Z$ 
\\ [0.5ex]
\hline
All                     &2\,191 &$ +12 \pm 0$    &$ -2 \pm 0$    &$+3 \pm 0$ \\
Defining                &262            &$  -1 \pm 1$      &$ -2 \pm 1$    &$+5 \pm 1$ \\
Non-defining    &1\,929 &$ +13 \pm 0$    &$ -2 \pm 0$    &$+3 \pm 0$ \\
Frame-fixed     &260            &$  -0 \pm 1$      &$ -2 \pm 1$    &$+5 \pm 1$ \\ [0.5ex]
\hline
\end{tabular}
\tablefoot{The unit is\uas.}
\end{table*}
Table~\ref{tab:GAEffect} reports the influence of the Galactic aberration effect on the {\it Gaia} DR1 reference frame.
We found no significant components except for $\Delta\epsilon_x$\,$\sim$\,0.01\mas, which would cause a displacement of the $Z$ axis and should possibly be  taken into consideration in the final {\it Gaia}-ICRF alignment.

\section{Global rotation between TGAS and {\it Tycho}-2 proper motions}    \label{sec:GlobalRotation}
The {\it Hipparcos} reference frame is estimated to rotate with respect to the {\it Gaia} DR1 reference frame at a rate of 0.24\masyr, and this rotation can be represented by a vector $\bm \omega \simeq (-0.126, +0.185, +0.076)^{\rm T}$\masyr\citep[][]{2016A&A...595A...4L}.
The {\it Tycho}-2 catalogue contains position and proper motion data for 2.5 million brightest stars, referring to the {\it Hipparcos} reference frame \citep{2000A&A...355L..27H}.
The joint {\it Tycho-Gaia} astrometric solution \citep[TGAS;][]{2015A&A...574A.115M} provides astrometric parameters for stars common in the {\it Tycho}-2 catalogue in the frame of the {\it Gaia} DR1 reference frame.
Therefore, in a global sense, the {\it Tycho}-2 proper motion system should differ from the TGAS proper motion system by a global rotation that may be not totally consistent with, but comparable to, $\bm \omega$ in the magnitude.
\citet{2016A&A...595A...4L} compared the TGAS positions and proper motions with the {\it Hipparcos} and {\it Tycho}-2 catalogues for individual sources, but did not consider the global difference between the proper motion systems.
This motivated us to perform  further analysis of the overall difference between the {\it Tycho}-2 and TGAS proper motion systems.

In {\it Gaia} DR1, all sources were treated as single stars.
Any astrometric effects due to the orbital motion in binaries or the perspective acceleration were ignored.
As a result, for the binaries, multiples, and suspected non-single systems, the nonlinear motion of the photo-centre may cause an instantaneous proper motion.
To avoid these distortions in the proper motion system, we only considered single stars in the {\it Tycho}-2 catalogue
(The flag ``posflg'' is set as $\sqcup$
\footnote{In the {\it Tycho}-2 catalogue,  the flag ``posflg'' indicates the type of {\it Tycho}-2 solution: $\sqcup\,=$\,normal treatment, ``D''\,=\,double star treatment, and ``P''\,=\,photo-centre treatment. })
and hence we obtained a sample of 1\,969\,315 single stars that are common to  TGAS and the {\it Tycho}-2 catalogues.

To exclude the effect of color-dependent errors in the proper motion systems \citep[][]{2016A&A...595A...1G}, we divided the stars into three groups: all stars, the O-B5 stars (young stars, age older than $3\,\times\,10^7\,\mathrm{yr}$ to reject the stars in the Gould belt),  and the K-M giants (luminosity class III).
These classification need further information on the MK spectral types and the luminosity classes, which is provided in the {\it Tycho}-2 Spectral Type Catalog \citep{2003AJ....125..359W}.
By cross-identification, we obtained a group of 5\,479 O--B5 stars and 32\,242 K--M giants.

The global rotation between TGAS and {\it Tycho}-2 can be determined by the least squares fit, using the following equations:
\begin{equation}\label{eq:rotation}
   \begin{array}{lll}
\Delta\mu_{\alpha ^*}   & = & -\omega^\prime_X\cos\alpha\sin\delta - \omega^\prime_Y\sin\alpha\sin\delta +
\omega^\prime_Z\cos\delta \\
\Delta\mu_{\delta}      & = &+\omega^\prime_X\sin\alpha                 - \omega^\prime_Y\cos\alpha,
   \end{array}
,\end{equation}
where $\mu_{\alpha ^*}=\mu_\alpha\cos\delta$.
The vector $\bm \omega^\prime =(\omega^\prime_X, \omega^\prime_Y, \omega^\prime_Z)^{\rm T}$ represents the spin between the two celestial reference frames, taken in the sense {\it Tycho}-2\,$-$\,TGAS.
The 2.6-$\sigma$ principle was introduced in the least squares to exclude the outlier proper motions.

From the results in Table~\ref{tab:GlobalSpin_tycho2}, we found no obvious global rotations for all and M--K giants, and a relatively large $\omega^\prime_Y$ component for the group of the O--B5 stars.
However, none of them has a magnitude exceeding 0.10\masyr.
The global rotation $\bm \omega^\prime$ in Table~\ref{tab:GlobalSpin_tycho2}, by directly comparing the stellar proper motions, is much smaller than the value (0.24\masyr), which was obtained by combining the stellar positions at J1991.25 with the quasar positions at J2015.0 \citep[][]{2016A&A...595A...4L}.
The {\it Tycho}-2 proper motions are derived by incorporating the original {\it Tycho}-1 catalogue \citep{1997ESASP1200.....E} with century-old ground-based catalogues, and thus contain systematic errors from old data, possibly responsible for the inconsistency.

Then we compared the TGAS proper motions with the revised {\it Hipparcos} proper motions, using the same method as for the {\it Tycho}-2 catalogue.
The revised {\it Hipparcos} catalogue \citep{2007A&A...474..653V} is a new reduction of the astrometric data of the {\it Hipparcos} mission with improved accuracies for astrometric parameters, compared with the original catalogue, and the reference frame remains the same as for the old {\it Hipparcos} catalogue.
Unlike the {\it Tycho}-2 proper motions, the revised {\it Hipparcos} proper motions contain no systematic errors from old ground-based catalogues.
Thus the comparison between TGAS and the revised {\it Hipparcos} proper motions were supposed to yield a reliable global rotation vector.
The results obtained from different samples (Table~\ref{tab:GlobalSpin_hip2}) yielded similar global rotation components to those in Table~\ref{tab:GlobalSpin_tycho2}, indicating that the systematic errors from the old data in {\it Tycho}-2 catalogue should not be responsible for the inconsistency between $\bm \omega$ and $\bm \omega^\prime$.

To illustrate possible causes, we need further investigations of the TGAS proper motions. In the following section, we analyzed the kinematics property of the Milky Way using the TGAS proper motion to deepen our understanding of the {\it Gaia} DR1 reference frame.

\begin{table*}[tp]
 \centering
\caption{\label{tab:GlobalSpin_tycho2}
Global Rotation between TGAS and {\it Tycho}-2 proper motions.}
\begin{tabular}{c c c c c c}
\hline \hline
Sample          &  $N$ (All / Outlier)          &$\omega^\prime_X$      &$\omega^\prime_Y$       &$\omega^\prime_Z$      &$\omega^\prime$        \\ [0.5ex]
\hline
All                     & 1\,969\,315 / 56\,349 &$+11 \pm   4$          &$+13 \pm   4$    &$+24 \pm   5$      &$29 \pm   8$ \\
O--B5 stars     & 5\,479 / 106                  &$ -28 \pm 48$          &$+79 \pm 48$    &$+8 \pm 58$ &$84 \pm 89$ \\
K--M giants     & 32\,242 / 882                 &$ -17 \pm 13$          &$+  3 \pm 13$    &$ -13 \pm 16$    &$21 \pm 24$ \\  [0.5ex]
\hline
\end{tabular}
\tablefoot{The unit is\uasyr.}
\end{table*}

\begin{table*}[tp]
 \centering
\caption{\label{tab:GlobalSpin_hip2}
Global rotation between TGAS and the revised {\it Hipparcos} proper motions.}
\begin{tabular}{c c c c c c}
\hline \hline
Sample          &  $N$ (All / Outlier)          &$\omega^\prime_X$      &$\omega^\prime_Y$       &$\omega^\prime_Z$      &$\omega^\prime$        \\[0.5ex]
\hline
All                     & 86\,849 / 2\,948              &$+ 8 \pm   7$           &$+10 \pm   7$    &$ -14 \pm   9$       &$19 \pm 13$ \\
O--B5 stars     & 1\,904 / 83                   &$+10 \pm 30$           &$+59 \pm 30$    &$ -34 \pm 37$       &$69 \pm 56$\\
K--M giants     & 13\,308 / 509                 &$ - 9 \pm 13$          &$+10 \pm 13$    &$ -  9 \pm 15$      &$16 \pm 24$ \\
[0.5ex] 
\hline
\end{tabular}
\tablefoot{The unit is\uasyr.}
\end{table*}

\section{Property of {\it Gaia} DR1 reference frame inferred from Galactic kinematics analysis}    \label{sec:KinematicsAnalysis}
\subsection{Analytic strategy}
The proper motion system of a catalogue not only reflects the inertial characteristic of the stellar reference frame, but also critically affects  the analysis on the kinematics and dynamics of the Galaxy \citep{2007ChA&A..31..296Z}.
The {\it Gaia} DR1 reference frame is an ideally rigid and inertial reference frame with respect to the extragalactic objects, so that the observed stellar proper motions in this reference frame are the sum of the following three terms except for the observational errors:
\begin{enumerate}
 \item the stellar peculiar motion;
 \item the solar peculiar motion (the origin of the reference frame is set on the solar system barycenter);
  \item the Galactic rotation.
\end{enumerate}
The stellar peculiar motion is always considered as an ellipsoidal velocity distribution and statistically treated as the random variables with zero mean.
If one uses the TGAS proper motions of a specific group of stars and performs the kinematical analysis of the Galaxy, the solution should be compatible with the results obtained by the physical theories or models to within uncertainties.

The sample of K-M giants, one of the old and well-relaxed populations of stars, is supposed to be a steady-state constituent of the Galaxy.
This  stellar ensemble should exhibit only the familiar in-plane galactic rotation (described by the Oort constants {\it A} and {\it B}) \citep{1993AJ....105..691M} in an ideal inertial reference frame.
The unexplained motion of stars should be attributed to possible problems of the current Galactic kinematical model or the {\it Gaia} DR1 reference frame.

\subsection{Materials and results}
To investigate the property of the reference frame, we adopt the Ogorodmikov--Milne model \citep{1935MNRAS..95..560M}, in which the stellar proper motion field in the solar neighborhood can be described by
\begin{equation}\label{eq:OMModel}
\left(
\begin{array}{c}
\mu_l \cos b    \\
\mu_b\\
\end{array} \right) = {\mathcal M}X,
\end{equation}
where $\mathcal M$ is a $2\,\times\,9$ matrix consisting of known trigonometric functions of the galactic coordinates of stars.
The vector $\bm{X}$ can be given by the following equation:
\begin{equation}\label{eq:9Parameters}
\bm{X}^{\rm T} = (S_1, S_2, S_3, D^-_{32}, D^-_{13}, D^-_{21}, D^+_{12}, D^+_{13}, D^+_{23} )^{\rm T}
,\end{equation}
where the unknowns can be divided into three components: the solar peculiar motion $(S_1, S_2, S_3)^{\rm T}$, the Galactic rigid rotation $(D^-_{32}, D^-_{13}, D^-_{21})^{\rm T}$, and the Galactic differential rotation $(D^+_{23}, D^+_{13},  D^+_{12})^{\rm T}$.
The rotation components $D^+_{12}$ and $D^-_{21}$ are equivalent to the Oort constants $A$ and $B$ in the two-dimensional case.

The initial sample of K-M giants is the same as that in Sect.~\ref{sec:GlobalRotation} (32\,242 K--M giants in Table~\ref{tab:GlobalSpin_tycho2}).
In the next step, we reject 89 objects with the negative parallax and 4\,425 with the relative parallax uncertainty larger than 30\%.
Then we set limits for the heliocentric distance $r$ and the galactic height $z$:
\begin{equation}\label{eq:DistributionFilter}
0.2\,\mathrm{kpc} \le r \le 1.0\,\mathrm{kpc},  \quad | z | \le 0.5\,\mathrm{kpc}
\end{equation}
The upper limit of $1\,\mathrm{kpc}$ for $r$ was set, since the Ogorodmikov--Milne model is the first order Taylor expansion near the Sun.
In the solar vicinity, the velocity dispersions may dominate the proper motion field rather than the systematic motion owing to the small distance,
so that a lower limit of $0.2\,\mathrm{kpc}$ was set for $r$ to avoid this effect.
A maximum of $0.5\,\mathrm{kpc}$ for $z$ was set to keep the stars in the thick disk.
As a result, we obtained a sample of 23\,612 K--M giants in this step.

\begin{figure}[hbtp]
  \centering
  \includegraphics[width=\columnwidth]{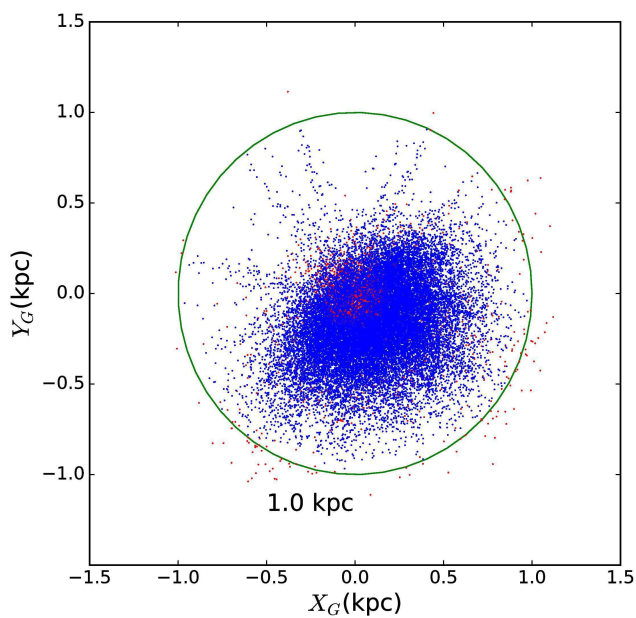}
  \caption[]{\label{fig:XY_mk}
    Distribution of the sample of all K--M III giants on $X_G$--$Y_G$ plane of Galactic coordinate. Blue points represent the stars accepted by Eq.~(\ref{eq:DistributionFilter}), while red points stand for the rejected ones.
  }
\end{figure}

Figure~\ref{fig:XY_mk} gives the sample distribution in $X_G$--$Y_G$ plane of the galactic coordinate system.
One can clearly see an inhomogeneous distribution, which may arise from the incompleteness of the spectral catalog and property of the scanning law in the {\it Gaia} observation.

To test the possible influence of the inhomogeneous distribution of the sample on determining the vector $X$, we set different lower limits for $r$ and obtained different samples. Then we applied the least squares to Eq.~(\ref{eq:OMModel}) with different star samples.
The results are given as solution A to D in Table~\ref{tab:KinematicResult} for samples with different lower limits of $r$.
The circular rotation velocity $V_0$ at the Sun is obtained by combining the rotational components $D_{21}^-$ and $D_{12}^+$ and adopting the solar distance to the Galactic center $R_0 = 8.34$\,kpc \citep{2014ApJ...783..130R}.
The determination of the solar peculiar motion and the Oort's constants $D_{12}^+$ and $D_{21}^-$ was robust, leading to a relatively stable value of $V_0,$ which is consistent with the result $240\,\pm\,8$\kms~ in \citet{2014ApJ...783..130R}.
The rotational components $D_{13}^-$ and $D_{13}^+$ around the $Y_G$ axis seemed to be non-zero, while the components $D_{32}^-$ and $D_{23}^+$ around the $X_G$ axis were estimated to be zero within the standard error in most cases.

We then applied a constraint $D_{32}^-=0, D_{23}^+=0$ in the least squares and the number of parameters to be determined in Eq.~(\ref{eq:9Parameters}) became seven. The results are shown as solution E in Table~\ref{tab:KinematicResult}. Only the result of sample with 0.2\,kpc\,$\le r \le$\,1.0\,kpc is given; other samples yield similar results.
Compared with the result solution A without any constraint (see Row~1 in Table~\ref{tab:KinematicResult}), we found very little difference and hence considered that the constraint is reasonable.
In other words, no obvious $D_{13}^-$ and $D_{13}^+$ components could be found from the TGAS K--M giant proper motions.
In contrast, the non-zero Galactic rotational components around the $Y_G$ axis might exist and the solution E gives 
\begin{equation}\label{eq:NonZeroComponent}
\begin{array}{c }
\omega_{Y_G}             = D_{13}^- = -0.38 \pm 0.15\,\mathrm{mas\,yr^{-1}}   \\
\omega^\prime_{Y_G} = D_{13}^+ = -0.29 \pm 0.19\,\mathrm{mas\,yr^{-1}}
\end{array}
.\end{equation}
\citet{2016A&A...595A...2G} point out that the {\it Gaia} DR1 parallax system have a systematic error of $\sim$0.3\,mas, which were validated by \citet{2041-8205-831-1-L6} and \citet{2016ApJ...832L..18J}, respectively.
To test the possible influence of the parallax systematic error on our results, we enlarged the parallax, i.e. by a factor of 110\% or 120\%, and performed the same least squares fits. The test we performed indicated that changes in parallax affected the determination of the solar peculiar motion and the Oort's constants but not the other Galactic rotational components such as $D_{13}^-$ and $D_{13}^+$.
 \begin{table*}
\caption{\label{tab:KinematicResult}
Result derived from the TGAS proper motions of K-M giants in the Galactic rectangular coordinates.
The first six rows correspond to the least squares of Eqs.~(\ref{eq:OMModel})$\,\sim\,$(\ref{eq:9Parameters}) while the last row is obtained with the extra constraint $D_{32}^-=D_{23}^+=0$.}
\begin{tabular}{ c c c c c c c c c c c c c}
\hline \hline
Solution                                        &
$r$\tablefootmark{a}                    &
$N$                                             &
$S_1$\tablefootmark{b}          &
$S_2$\tablefootmark{b}                  &
$S_3$\tablefootmark{b}          &
$D^-_{32}$\tablefootmark{c}     &
$D^-_{13}$\tablefootmark{c}     &
$D^-_{21}$\tablefootmark{c}     &
$D^+_{12}$\tablefootmark{c}     &
$D^+_{13}$\tablefootmark{c}     &
$D^+_{23}$\tablefootmark{c}     &
$V_0$\tablefootmark{b} \\[0.5ex]
\hline

\multirow{2}{*}{A}              &\multirow{2}{*}{[0.2, 1.0]}    &23\,612
&$9.9 $    &$20.7 $   &$6.5 $    &$1.2 $    &$-1.6 $   &$-13.2 $      &$15.4 $   &$-1.6 $   &$0.7 $    &$238.6 $ \\
                                        &                                       &(1\,067)\tablefootmark{d}
&$\pm 0.3$     &$\pm 0.3$     &$\pm 0.3$     &$\pm 0.8$     &$\pm 0.7$     &$\pm 0.7$     &$\pm 0.9$     &$\pm 0.9$     &$\pm 0.9$     &$\pm 9.7$ \\

\multirow{2}{*}{B}              &\multirow{2}{*}{[0.3, 1.0]}    &18\,685
&$10.2 $   &$21.4 $   &$6.6 $    &$0.9 $    &$-1.6 $   &$-12.1 $      &$15.1 $   &$-0.7 $   &$-0.3 $   &$227.0 $ \\
                                        &                                       &(832)
&$\pm 0.4$     &$\pm 0.3$     &$\pm 0.4$     &$\pm 0.7$     &$\pm 0.7$     &$\pm 0.7$     &$\pm 0.9$     &$\pm 0.9$     &$\pm 0.9$     &$\pm 9.4$ \\

\multirow{2}{*}{C}              &\multirow{2}{*}{[0.4, 1.0]}    &12\,533
&$10.6 $   &$20.5 $   &$6.3 $    &$0.4 $    &$-1.9 $   &$-11.6 $      &$14.7 $   &$-1.7 $   &$0.6 $    &$219.9 $ \\
                                        &                                       &(556)
&$\pm 0.5 $     &$\pm 0.4$     &$\pm 0.5$     &$\pm 0.8$     &$\pm 0.8$     &$\pm 0.8$     &$\pm 0.9$     &$\pm 0.9$     &$\pm 0.9$     &$\pm 10.0$  \\                      
                        
\multirow{2}{*}{D}              &\multirow{2}{*}{[0.5, 1.0]}    &7\,233
&$10.4 $   &$20.1 $   &$7.0 $    &$-0.1 $   &$0.0 $    &$-12.3 $      &$15.7 $   &$0.7 $    &$0.3 $    &$233.7 $ \\
                                        &                                       &(321)
&$\pm 0.7$     &$\pm 0.6$     &$\pm 0.6$     &$\pm 1.0$     &$\pm 0.9$     &$\pm 0.9$     &$\pm 1.1$     &$\pm 1.1$     &$\pm 1.1$     &$\pm 11.7$        \\

\hline

\multirow{2}{*}{E}              &\multirow{2}{*}{[0.2, 1.0]}    &23\,612
&$9.9 $    &$20.7 $   &$6.8 $    &0             &$-1.8 $   &$-13.3 $      &$15.4 $   &$-1.4 $   &0       &$239.6 $ \\
                                        &                                       &(1\,064)
&$\pm 0.3 $     &$\pm 0.3 $     &$\pm 0.2 $     &                       &$\pm 0.7$     &$\pm 0.7$     &$\pm 0.9 $     &$\pm 0.9 $     &               &$\pm 9.7 $ \\  [0.5ex]

\hline
\end{tabular} \\
\tablefoot{
\tablefoottext{a}{Upper and lower limits of the heliocentric distance $r$. Units: kpc.} \\
\tablefoottext{b}{Units: \kms.} \\
\tablefoottext{c}{Units: \kmskpc.} \\
\tablefoottext{d}{Outliers due to the filter of 2.6-$\sigma$ principle in the least squares fit.} \\ }
\end{table*}
\subsection{Possible residual rotation of the {\it Gaia} DR1 reference frame}
Assuming that the Galactic rotations around the $X_G$ and $Z_G$ axis were well determined and taking only the two non-zero components $\omega_{Y_G}$ and $\omega^\prime_{Y_G}$ into consideration, the possible residual rotation of the reference frame in equatorial coordinates can be calculated:
\begin{equation}\label{eq:RigidRotation}
\left(
\begin{array}{c}
\omega_1                \\
\omega_2                \\
\omega_3
\end{array} \right) = \mathcal{N}_{J2000}^{\rm T}
\left(
\begin{array}{c}
0                               \\
\omega_{Y_G}            \\
0
\end{array} \right)
=
\left(
\begin{array}{c}
 -0.19          \\
+0.17           \\
 -0.28
\end{array} \right)\,\mathrm{mas\,yr^{-1}}
\end{equation}
\begin{equation}\label{eq:DifferentialRotation}
\left(
\begin{array}{c}
\omega^\prime_1         \\
\omega^\prime_2         \\
\omega^\prime_3
\end{array} \right) = \mathcal{N}_{J2000}^{\rm T}
\left(
\begin{array}{c}
0                               \\
\omega^\prime_{Y_G}             \\
0
\end{array} \right)
=
\left(
\begin{array}{c}
 -0.14          \\
+0.13           \\
 -0.22
\end{array} \right)\,\mathrm{mas\,yr^{-1}}
,\end{equation}
where $\mathcal{N}_{J2000}$ is the commonly used equatorial-to-Galactic transformation matrix, i.e. adopted in \citet{1997ESASP1200.....E}.

However, the determination of unknowns is limited by the sample.
The sample of all the available K--M giants yields the inhomogeneous and anisotropic distribution on the $X_G$--$Y_G$ plane (as seen from Fig.~\ref{fig:XY_mk}).
Additionally, the {\it Tycho}-2 Catalogue is complete at a magnitude of about V$\sim$11.0, therefore the sample we obtained is a subsample of bright K--M giants.
Table~\ref{tab:OM_coeff} reports the correlation coefficients in solution E.
We  find a strong correlation between the solar peculiar motion and the Galactic rotation (i.e. $S_1$ and $D_{21}^-$), which  might be caused by the limitation of the sample.
Hence the residual rotations that were obtained here require further investigations.
 \begin{table}
 \centering
\caption{\label{tab:OM_coeff}
Correlation coefficients of seven unknowns whose estimations are reported in the last row of Table~\ref{tab:KinematicResult}.}
\begin{tabular}{l r r r r r r r  }
\hline \hline
 &$S_1$         &$S_2$          &$S_3$  &$D^-_{13}$     &$D^-_{21}$     &$D^+_{12}$      &$D^+_{13}$      \\ [0.5ex]
\hline

$S_1$           &1      &+0.0   &$- 0.1$        &$- 0.3$                &+0.5           &$ -0.4$                    &$-0.3$         \\
$S_2$           &               &1      &$+0.0$ &$- 0.0$                &+0.3           &+0.3                    &+0.1    \\
$S_3$           &               &               &1      &$-0.3$         &+0.0           &+0.1                    &+0.2    \\
$D^-_{13}$      &               &               &               &1              &$-0.2$         &+0.1                    &+0.2 \\
$D^-_{21}$      &               &               &               &                       &1              &$-0.1$                  &$-0.1$ \\
$D^+_{12}$      &               &               &               &                       &                       &1                       &+0.1  \\
$D^+_{13}$      &               &               &               &                       &                       &                                &1 \\    [0.5ex]

\hline
\end{tabular}
 \end{table}
\section{Conclusions} \label{sec:conclusions}
In this paper, we present a detailed analysis of the overall property of the {\it Gaia} DR1 reference frame.
Initially, we validated the declination bias of $\sim$$-$0.1\,mas in {\it Gaia} auxiliary quasar solution with respect to the ICRF2 and find that the systematic declination bias exists in both northern and southern hemispheres. 
The Galactic aberration effect is thought to produce an offset $\sim$0.01\,mas of the $Z$ axis over the J2000.0-J20015.0 period, which should be taken into consideration in the future {\it Gaia}-ICRF link.
All the tests indicate that {\it Gaia} DR1 reference frame is well aligned to ICRF2 and a quasi-inertial reference frame with respect to extragalactic objects.

To compare the TGAS with {\it Tycho}-2 proper motion systems, we picked out single stars from the {\it Tycho}-2 catalogue, divided them into three groups according to the spectral types and luminosity classes, and determined the global rotation of the {\it Hipparcos} reference frame relative to the {\it Gaia} DR1 reference frame by a least squares fit.
Although different samples give different values for global rotation (Tables~\ref{tab:GlobalSpin_tycho2} and \ref{tab:GlobalSpin_hip2}), the magnitude of global rotation obtained is much smaller than 0.24\masyr~in \citet{2016A&A...595A...4L}.

The kinematical analysis of the TGAS K-M giant proper motions give consistent results of the solar peculiar motion and Oort constants to those in literature.
But we found possible non-zero components of Galactic rotation  (the rigid rotation component $\omega_{Y_G}$ and the differential rotation component $\omega^\prime_{Y_G}$), which were non-negligible in our solution.
This result indicates a possible residual rotation in the {\it Gaia} DR1 reference frame or problems in the current Galactic kinematical model.
However, owing to the limitation of the sample domain, the result is not robust at present.
Here we call attention to the possible residual rotation presented in this work, which should be carefully examined in later studies and the future {\it Gaia} data release.

\begin{acknowledgements}
The authors are grateful to Dr Zacharias, who provided very useful comments and suggestions to improve the manuscript. 
This research is funded by the National Natural Science Foundation of China (NSFC) under grant No. 11473013.
This work  made use of data from the European Space Agency (ESA)
mission {\it Gaia} (\url{http://www.cosmos.esa.int/gaia}), processed by
the {\it Gaia} Data Processing and Analysis Consortium (DPAC,
\url{http://www.cosmos.esa.int/web/gaia/dpac/consortium}). Funding
for the DPAC was provided by national institutions, in particular
the institutions participating in the {\it Gaia} Multilateral Agreement.
This research  made use of the SIMBAD database, operated at CDS, Strasbourg, France.
\end{acknowledgements}

%
\bibliographystyle{aa} 
\bibliography{GaiaDR1_CRF} 
%
\end{document}